\documentclass[twocolumn,showpacs,amsmath,amssymb,prd]{revtex4}
             
\usepackage{graphicx}
\usepackage{dcolumn}
\usepackage{bm}
\usepackage{mathrsfs,amsmath}

\def\beq{\begin{equation}}
\def\enq{\end{equation}}  
\def\ba{\begin{eqnarray}}
\def\ea{\end{eqnarray}}
\def\<{\langle}
\def\>{\rangle}

\begin{document}

\title{Future CMB cosmological constraints in a dark coupled universe} \author{Matteo Martinelli$^1$, Laura L\'opez Honorez$^{2}$, Alessandro Melchiorri$^1$ and Olga Mena$^3$} 
\affiliation{$^1$Physics Department and INFN, Universita' di Roma ``La Sapienza'', Ple Aldo Moro 2, 00185, Rome, Italy.}
 \affiliation{$^2$Physics Department and IFT, UAM, 28049 Cantoblanco, Madrid, Spain and\\
 Service de Physique Th\'eorique, ULB, 1050 Brussels, Belgium}
 \affiliation{$^3$IFIC-CSIC and Universidad de Valencia, Valencia, Spain}

\begin{abstract}
Cosmic Microwave Background satellite missions as the on-going Planck
experiment are expected to provide the strongest constraints on a 
wide set of cosmological parameters. Those constraints, however, 
could be weakened when the assumption of a cosmological constant as the
dark energy component is removed. Here we show that it will indeed be 
the case when there exists a coupling among the dark energy and the 
dark matter fluids. In particular, the expected errors on key 
parameters as the cold dark matter density and the angular diameter distance at decoupling are 
significantly larger when a dark coupling is introduced. 
We show that it will be the case also for future satellite missions 
as EPIC, unless CMB lensing extraction is performed.

\end{abstract}


\date{\today}
\maketitle

\section{Introduction}

Current cosmological measurements point to a \emph{flat} universe
whose mass-energy includes $5\%$ ordinary matter and $22\%$
non-baryonic dark matter, but is dominated by the \emph{dark energy}
component, identified as the engine for the accelerated
expansion~\cite{wmap5,Wood-VaseySN07,TegmarkLRGDR4,PercivalLRG,reid,percival,wmap7}. The
most economical description of the cosmological measurements
attributes the dark energy to a Cosmological Constant (CC) in
Einstein's equations, representing an invariable vacuum energy
density. The equation of state of the dark energy component $w$ in the
CC case is constant and $w=-1$. However, from the quantum field
approach, the bare prediction for the current vacuum energy density is
$\sim 120$ orders of magnitude larger than the observed value. This
situation is the so-called CC problem. In addition, there is no
proposal which explains naturally why the matter and the vacuum energy
densities give similar contributions to the universe's energy budget
at this moment in the cosmic history. This is the so-called \emph{why
  now?} problem, and a possible way to alleviate it is to assume a
time varying, dynamical fluid. The quintessence option consists on a
cosmic scalar field $\phi$ (called \emph{quintessence} itself) which
changes with time and varies across space, and it is slowly
approaching its ground state.
Also, the quintessence equation of state is generally not constant
through cosmic
time~\cite{RatraPeebles88a,RatraPeebles88b,Wetterich95,Caldwell98,
  Quint99,Wang00}. In principle, the quintessence field may couple to
the other fields. In practice, observations strongly constrain the
couplings to ordinary matter~\cite{carroll}. However, interactions
within the dark sectors, i.e. between dark matter and dark energy, are
still allowed. This could change significantly the universe
 and the density perturbation evolution, the latter being seeds for
structure formation. For models equivalent to the one studied here,
see {\it e.g.} Refs.~\cite{amendola,Valiviita:2008iv,He:2008si,Jackson:2009mz,Gavela:2009cy,CalderaCabral:2009ja,Valiviita:2009nu,Majerotto:2009np}. \\

In this paper we investigate how allowing for a feasible interacting dark matter and dark energy model will affect the cosmological
constraints expected from future CMB experiments. The Planck satellite mission, for example, is expected to provide high quality
constraints on several key parameters (see
e.g. \cite{Perotto:2006rj}). However, those forecasts are usually
performed under  the assumption that the dark energy component is either a cosmological constant or a fluid with constant, redshift 
independent equation of state $w=P/\rho$. It is therefore timely to investigate if the assumption of a more elaborate dark energy component with a coupling with the dark matter could have an impact on these constraints.
Here we indeed focus on the future CMB data constraints on interacting
dark matter-dark energy models, exploiting, in particular, the gravitational CMB lensing signal. The structure of the paper is as follows. Section \ref{sec:seci} presents the background and the linear perturbations of the interacting dark matter-dark energy model explored here. Sections \ref{sec:secii} and \ref{sec:seciii} describe the CMB lensing extraction method and the future CMB data simulation used in our numerical analysis, respectively. We present our results in Sec.~\ref{sec:seciiii} and draw our conclusions in Sec.~\ref{sec:seciiiii}. 

\section{Preliminaries}
\label{sec:seci}
At the level of the background evolution equations, one can generally introduce a coupling between the dark matter and dark energy sectors as follows:
\begin{eqnarray}
   \label{eq:EOMm}
  \dot{\bar \rho}_{dm}+ 3  \mathcal{H} \bar\rho_{dm} &=&a \bar Q\,,\\
\label{eq:EOMe}
 \dot{\bar \rho}_{de}+ 3 \mathcal{H} \bar \rho_{de}(1+ w)&=&-a \bar Q\,,
\end{eqnarray}
where the bars denotes background quantities,  $\bar \rho_{dm} (\bar
\rho_{de})$ refers to  the dark matter (dark energy) energy density, the
dot indicates derivative with respect to conformal time $d\tau = dt/a$
and $w=\bar P_{de}/ \bar\rho_{de}$ is the dark-energy equation of state ($P$
denotes the pressure). We take $\mathcal{H}= {\dot a}/a$ as the
background expansion rate. We work in the context of a
Friedman-Robertson-Walker (FRW) metric, assuming a flat universe and
pressureless dark matter $w_{dm} = \bar P_{dm}/\bar \rho_{dm}=0$. $\bar Q$
encodes the dark coupling and drives the background energy exchange between dark
matter and dark energy. In order to deduce the evolution of the
background as well as the density and velocity perturbations in
coupled models, we need an expression for the energy transfer at the
level of of the stress-energy tensor: 
\begin{eqnarray}
\nabla_\mu T^{\mu\nu}_{(a)} =Q^\nu_{(a)}~,
\label{eq:conservDMDE}
\end{eqnarray}
where $a= dm, de$, $T^{\mu\nu}$ refers to the energy-momentum tensor and $Q^\nu_{(a)}$  is the energy-momentum transfer between the dark matter and dark energy fluids. We consider  
\begin{equation}
  Q_\nu^{(dm)}= \xi H \rho_{de} u_{\nu}^{(dm)}=-Q_\nu^{(de)} ~,
\label{eq:coupl}
\end{equation}
where $\xi$ is a dimensionless coupling (considered constant, as well
as $w$, in the
present analysis). $H$ and $\rho_{de}$ refer to the total
expansion rate and dark energy density, background plus perturbation, 
{\it i.e} $H={\cal H}/a+\delta H$ and $\rho_{de}=\bar
\rho_{de}+\delta \rho_{de}$ respectively. In the previous Eqs.~(\ref{eq:EOMm})
and~(\ref{eq:EOMe}), $\bar Q$ corresponds to $ \xi {\cal H}
\bar\rho_{de}/a$, see our choice of coupling in Eq.~(\ref{eq:coupl}).
Notice from Eq.~(\ref{eq:coupl}) that $Q_\nu^{(a)}$ has been chosen parallel
to the dark matter four velocity $u_{\nu}^{(dm)}$, in order to avoid
momentum transfer in the rest frame of the dark
matter component~\cite{Valiviita:2008iv}. For this choice of energy exchange
$Q_\nu^{(a)}$, positive (negative) values of the coupling $\xi$ will lead to
lower (higher) dark matter energy densities in the past than in the
uncoupled $\xi=0$ case. In the following, we restrict ourselves here
to negative couplings and $w > −1$, which avoids instability problems in 
the dark energy perturbation equations, see Ref.~\cite{Gavela:2009cy}.

The interacting model given by Eq.~(\ref{eq:coupl}) has already been
previously explored under several assumptions, see Refs.~\cite{Valiviita:2008iv,He:2008si,Jackson:2009mz,Gavela:2009cy}. 
In those works, the linear perturbation analysis did not include
perturbation of the expansion rate $\delta H$. Let us mention that the former
is included in the numerical analysis presented here. The latter is
quite relevant for the correct treatment  of
gauge invariant perturbation but it does not affect much the physical
results. Details of the  complete linear perturbation analysis will be presented in Ref.~\cite{Gavela:2010} including the specification of  the initial
conditions which have been chosen adiabatic for all the components
except for the dark energy fluid, see Ref.~\cite{Gavela:2010}. For the numerical analysis presented here, we have modified the publicly available CAMB code~\cite{camb}, taking into account the presence of the dark coupling in both the background and the linear perturbation equations.

\section{Lensing extraction}
\label{sec:secii}
The analysis presented here includes, in addition to the primary CMB anisotropy angular power spectrum, the information from CMB lensing. Gravitational CMB lensing, as already shown (see e.g. \cite{Perotto:2006rj,calabrese}, can improve significantly the CMB constraints on several cosmological parameters, since it is
strongly connected with the growth of perturbations and gravitational
potentials at redshifts $z < 1$ and therefore, it can break important 
degeneracies.
The lensing deflection field $d$ can be related to the lensing potential $\phi$ as $d=\nabla\phi$~\cite{hirata:2003}. In harmonic space, the deflection and lensing potential multipoles follows 
\begin{equation}
d_l^m=-i\sqrt{l(l+1)}\phi_l^m,
\label{eq:defd}
\end{equation}
\noindent and therefore, the power spectra $C^{dd}_l\equiv\left\langle d_l^m d_l^{m*}\right\rangle$ and 
$C_l^{\phi\phi}\equiv\left\langle\phi_l^m\phi_l^{m*}\right\rangle$ are related through
\begin{equation}
C_l^{dd}=l(l+1)C_l^{\phi\phi}.
\end{equation}

Lensing introduces a correlation between different CMB multipoles (that otherwise would be fully uncorrelated) through the relation
\begin{equation}
\left\langle a_l^m b_{l'}^{m'}\right\rangle=(-1)^m\delta_m^{m'}\delta_l^{l'}C_l^{ab}+
\sum_{LM}{\Xi^{mm'M}_{l\ l'\ L}\phi^M_L}~,
\label{eq:lens_corr}
\end{equation}
\noindent where $a$ and $b$ are the ${T,E,B}$ modes and $\Xi$ is a linear combination of the unlensed
power spectra $\tilde{C}_l^{ab}$ (see \cite{lensextr} for details).\\
In order to obtain the deflection power spectrum from the observed $C_l^{ab}$, we have to invert Eq.~(\ref{eq:lens_corr}), defining a quadratic estimator for the deflection field given by
\begin{equation}
d(a,b)_L^M=n_L^{ab}\sum_{ll'mm'}W(a,b)_{l\ l'\ L}^{mm'M}a^m_lb^{m'}_{l'}~,
\label{eq:estimator}
\end{equation}
\noindent where $n_L^{ab}$ is a normalization factor needed to construct an unbiased estimator ($d(a,b)$ must satisfy Eq.~(\ref{eq:defd})).
This estimator has a variance:
\begin{equation}
 \langle d(a,b)_L^{M*} d(a',b')_{L'}^{M'}\rangle\equiv \delta_{L}^{L'}\delta^{M'}_{M}(C_L^{dd}+N_L^{aa'bb'})
\end{equation}
that depends on the choice of the weighting factor $W$ 
and leads to a noise $N_L^{aa'bb'}$ on the deflection power spectrum $C_L^{dd}$ obtained through this method. In the next section we  describe the method followed to extract the lensing noise. \\

\section{Future CMB data analysis}
\label{sec:seciii}

We evaluate the achievable constraints on the coupling parameter $\xi$ by a 
COSMOMC analysis of future mock CMB datasets.
The analysis method we adopt here is based on the
publicly available Markov Chain Monte Carlo package \texttt{cosmomc}
\cite{Lewis:2002ah} with a convergence diagnostic using the
Gelman and Rubin statistics. We sample the following
seven-dimensional set of cosmological parameters, adopting flat priors
on them: the baryon and cold dark matter densities $\Omega_{b}h^2$ and
$\Omega_{c}h^2$, the ratio of the sound horizon to the angular diameter
distance at decoupling $\theta_s$, the scalar spectral index $n_s$,
the overall normalization of the spectrum $A_s$ at $k=0.002$ {\rm Mpc}$^{-1}$,
the optical depth to reionization $\tau$, and, finally, the coupling parameter $\xi$. \\

We create full mock CMB datasets (temperature, E--polarization mode and lensing
deflection field) with noise properties consistent with Planck~\cite{:2006uk} and EPIC~\cite{EPIC} experiments, see Tab.~\ref{tab:exp} for their specifications. The fiducial model is chosen to be the best-fit from the WMAP analysis of 
Ref.~\cite{wmap5} with $\Omega_{b}h^2=0.0227$, $\Omega_{c}h^2= 0.113$, $n_s=0.963$, $\tau=0.09$ and $\xi=0$, fixing $w=-0.9$ for our numerical calculations.\\

\begin{table}[!htb]
\begin{center}
\begin{tabular}{rccc}
Experiment & Channel & FWHM & $\Delta T/T$ \\
\hline
Planck & 70 & 14' & 4.7\\
\phantom{Planck} & 100 & 10' & 2.5\\
\phantom{Planck} & 143 & 7.1'& 2.2\\

$f_{sky}=0.85$\\
\hline
EPIC & 70 & 12' & 0.05\\
\phantom{EPIC} & 100 & 8.4' & 0.05\\
\phantom{EPIC} & 150 & 5.6' & 0.06 \\
$f_{sky}=0.85$ & & &\\
\hline
\end{tabular}
\caption{Planck and EPIC experimental specifications. Channel frequency is given in GHz, FWHM (Full-Width at Half-Maximum) in arc-minutes, and the temperature sensitivity per pixel in $\mu K/K$. The polarization sensitivity is $\Delta E/E=\Delta B/B= \sqrt{2}\Delta T/T$.}

\label{tab:exp}
\end{center}
\end{table}

We consider for each channel a detector noise of $w^{-1} =
(\theta\sigma)^2$, where $\theta$ is the FWHM (Full-Width at
Half-Maximum) of the beam assuming a Gaussian profile and $\sigma$ is
the temperature sensitivity $\Delta T/T$ (see Tab.~\ref{tab:exp} for the polarization sensitivity). We therefore add to each $C_\ell$ fiducial spectra a noise spectrum given by:
\begin{equation}
N_\ell = w^{-1}\exp(l(l+1)/l_b^2) \, ,
\end{equation}
where $l_b$ is given by $l_b \equiv \sqrt{8\ln2}/\theta$.\\

In this work, we use the method presented in \cite{lensextr} to
construct the weighting factor $W$ of Eq.~(\ref{eq:estimator}). In
that paper, the authors choose $W$ to be a function of the power
spectra $C_l^{ab}$, which include both CMB lensing and primary
anisotropy contributions. This choice leads to five quadratic
estimators, with $ab={TT,TE,EE,EB,TB}$; the $BB$ case is excluded
because the method of Ref.~\cite{lensextr} is only valid when 
the lensing contribution is negligible compared to the primary anisotropy, 
assumption that fails for the B modes in the case of Planck. 
In the case of EPIC we have decided to neglect the $BB$ channel 
since it may be contaminated by unknown foregrounds and/or 
it may be used for foregrounds removal. 
The results presented here for the EPIC mission have therefore 
to be considered as conservative. The five quadratic estimators 
can be combined into a minimum variance estimator which provides 
the noise on the power spectrum of the deflection field $C_l^{dd}$ 
\begin{equation}
N_l^{dd}=\frac{1}{\sum_{aa'bb'}{(N_l^{aba'b'})^{-1}}}~.
\end{equation}
We compute the minimum variance lensing noise for both Planck and EPIC experiments by means of a publicly available routine~\cite{lensnoise}. 

The datasets (which include the lensing deflection power spectrum) are analyzed with a full-sky exact likelihood routine ~\cite{lensnoise}.

\section{Results}
\label{sec:seciiii}
\begin{table}[!htb]
\begin{center}
\begin{tabular}{r|c|c|c|c}
Parameter & Planck & Planck Lens & EPIC &EPIC Lens \\
\hline
$\Delta{(\Omega_bh^2)}$ & $0.00015$ & $0.00012$ & $0.00004$ & $0.00003$\\
$\Delta{(\Omega_ch^2)}$ & $0.0297$ & $0.0296$ & $0.0295$ & $0.016$\\
$\Delta{(\theta_s)}$    & $0.00229$ & $0.00215$& $0.00216$ & $0.00102$\\
$\Delta{(\tau)}$        & $0.0047$ & $0.0041$& $0.0022$ & $0.0021$\\
$\Delta{(n_s)}$         & $0.0037$ & $0.0029$& $0.0018$ & $0.0013$\\
$\Delta{(\log[10^{10} A_s])}$ & $0.016$ & $0.012$& $0.005$ & $0.004$\\
$\Delta{(H_0)}$         & $2.34$ & $2.16$& $2.17$ & $1.33$\\
$\Delta{(\Omega_\Lambda)}$         & $0.064$ & $0.061$& $0.062$ & $0.038$\\
$\xi$ & $>-0.59$ & $>-0.54$& $>-0.56$ & $>-0.34$\\
\hline
\end{tabular}
\caption{$68 \%$ c.l. errors on cosmological parameters. Upper limits on $\xi$ are $95\%$ c.l. constraints.}
\label{tab:results}
\end{center}
\end{table}

\begin{table}[!htb]
\begin{center}
\begin{tabular}{r|c|c}
Parameter  & Planck Lens& CMBPOL Lens\\
\hline
$\Delta(\Omega_b h^2)$ & 0.00013 & 0.00003 \\ 
$\Delta(\Omega_c h^2)$ & 0.0010 & 0.0003 \\ 
$\Delta(\theta_s)$ & 0.00026 & 0.00005 \\ 
$\Delta(\tau)$ & 0.0042 & 0.0022 \\ 
$\Delta(n_s)$ & 0.0031 & 0.0014 \\ 
$\Delta(log[10^{10} As])$ & 0.013 & 0.005 \\ 
$\Delta{(H_0)}$         & $0.53$ & $0.12$\\
$\Delta{(\Omega_\Lambda)}$& $0.005$ & $0.001$ \\
\hline
\end{tabular}
\end{center}
\caption{$68 \%$ c.l. errors on cosmological parameters from Planck and EPIC with lensing extraction in the standard non interacting case ($\xi=0$).}
\label{tab:tabstandard}
\end{table}

Table~\ref{tab:results} summarizes the errors from Planck and EPIC future data on the main cosmological parameters when a coupling $\xi$ among the dark energy and dark matter fluids is introduced in the model. Notice that the errors on the cosmological parameters which are degenerate with $\xi$ are larger than the errors that one would get for these parameters within a standard cosmology, where $\xi=0$ (see Tab.~\ref{tab:tabstandard}). In particular, in the case of $\Omega_ch^2$, $H_0$, $\Omega_{\Lambda}$ and $\theta_s$, the errors we obtain here are one order of magnitude larger than the ones obtained with $\xi=0$. For comparison, we show in Tab.~\ref{tab:results} the parameter constraints both with and without lensing extraction. The cosmological parameter constraints from EPIC mock data are stronger than those coming from future Planck data when the CMB lensing signal is exploited. The reason for that is because the EPIC experiment is expected to drastically reduce the noise in the CMB lensing extraction.

\begin{figure}[h!]
\begin{center}
\hspace*{-1cm}  
\begin{tabular}{cc}
\includegraphics[width=8cm]{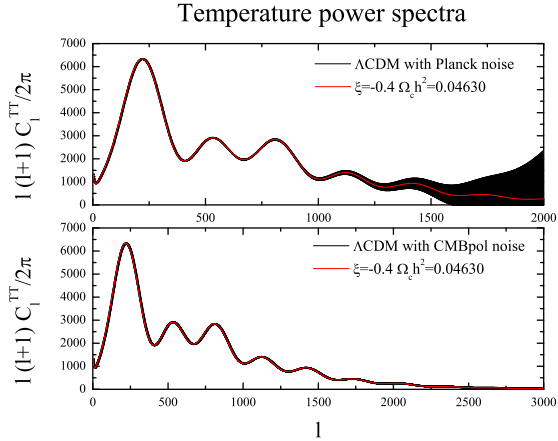}& 
 \end{tabular}
  \caption{Temperature power spectrum signal plus noise for Planck (top panel) and EPIC (bottom panel) experiments. The black curve depicts the $\Lambda$CDM model with $\Omega_{c}h^2= 0.113$. The red curve illustrates a coupled model allowed by Planck data, with $\xi=-0.4$ and $\Omega_{c}h^2= 0.0463$.} 
\label{fig:specnoise}
\end{center} 
\end{figure}

\begin{figure}[h!]
\begin{center}
\hspace*{-1cm}  
\begin{tabular}{cc}
\includegraphics[width=8cm]{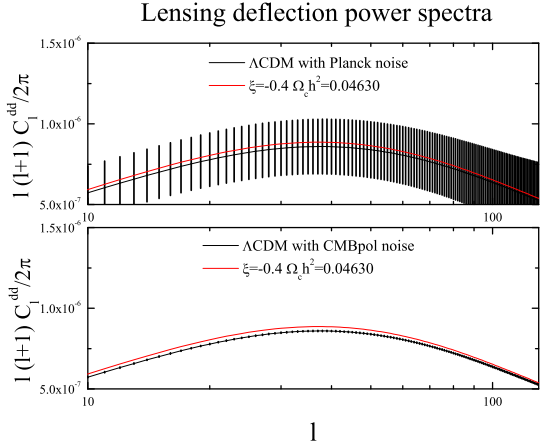}& 
 \end{tabular}
  \caption{Lensing deflection power spectrum signal plus noise for Planck (top panel) and EPIC (bottom panel) experiments. The black curve depicts the $\Lambda$CDM model with $\Omega_{c}h^2= 0.113$. The red curve illustrates a coupled model allowed by Planck data, with $\xi=-0.4$ and $\Omega_{c}h^2= 0.0463$.} 
  \label{fig:lensnoise}
\end{center} 
\end{figure}

Figures (\ref{fig:specnoise}) and (\ref{fig:lensnoise}) illustrate the CMB temperature and lensing deflection spectra plus noise from Planck and EPIC experiments assuming two possible cosmologies: a $\Lambda$CDM universe with $\Omega_{\rm c}h^2= 0.113$ and a coupled model with $\xi=-0.4$ and $\Omega_{\rm c}h^2= 0.0463$. From measurements of the CMB unlensed temperature spectrum (see Fig.~(\ref{fig:specnoise})) , these two models will be degenerate, since they have identical spectra, albeit with 
 different cold dark matter densities, since a more negative coupling can be compensated with a lower $\Omega_{c}h^2$. As already pointed out in Ref.~\cite{Gavela:2009cy}, in a universe with a negative coupling $\xi$, the matter content in the past can be higher than in the standard $\Lambda$CDM scenario due to an
extra contribution proportional to the dark energy component, and therefore 
$\Omega_ch^2$ is strongly correlated with the coupling $\xi$.

Notice from Fig.~(\ref{fig:specnoise}) that neither Planck nor EPIC data will be able to distinguish among coupled and uncoupled models using only primary CMB anisotropy data. However, these two models predict distinguishable lensing potential spectra, see Fig.~(\ref{fig:lensnoise}), and, while the Planck experiment will not have enough sensitivity to distinguish coupled versus uncoupled models, the EPIC experiment, with a greatly reduced noise on the CMB lensing extraction, will be able to test coupled models.

\begin{figure}[h!]
\begin{center}
\hspace*{-1cm}  
\begin{tabular}{cc}
\includegraphics[width=9cm]{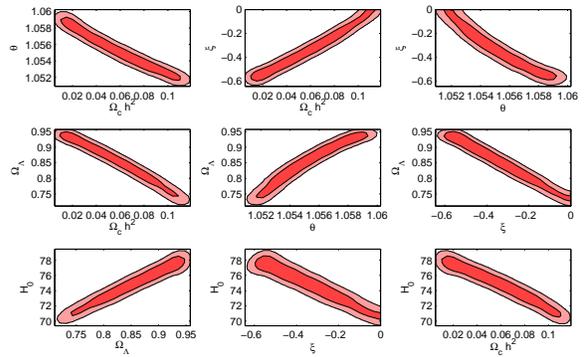}\\
 \end{tabular}
   \caption{The panels show the $68 \%$ and $95 \%$ confidence level contours 
combining the five most correlated parameters ($\Omega_ch^2$, $\theta_s$,
 $H_0$, $\Omega_\Lambda$ and $\xi$) arising from a fit to mock Planck data  without including lensing extraction in the analysis.}
  \label{fig:degeneracyPno}
\end{center} 
\end{figure}

\begin{figure}[h!]
\begin{center}
\hspace*{-1cm}  
\begin{tabular}{cc}
\includegraphics[width=9cm]{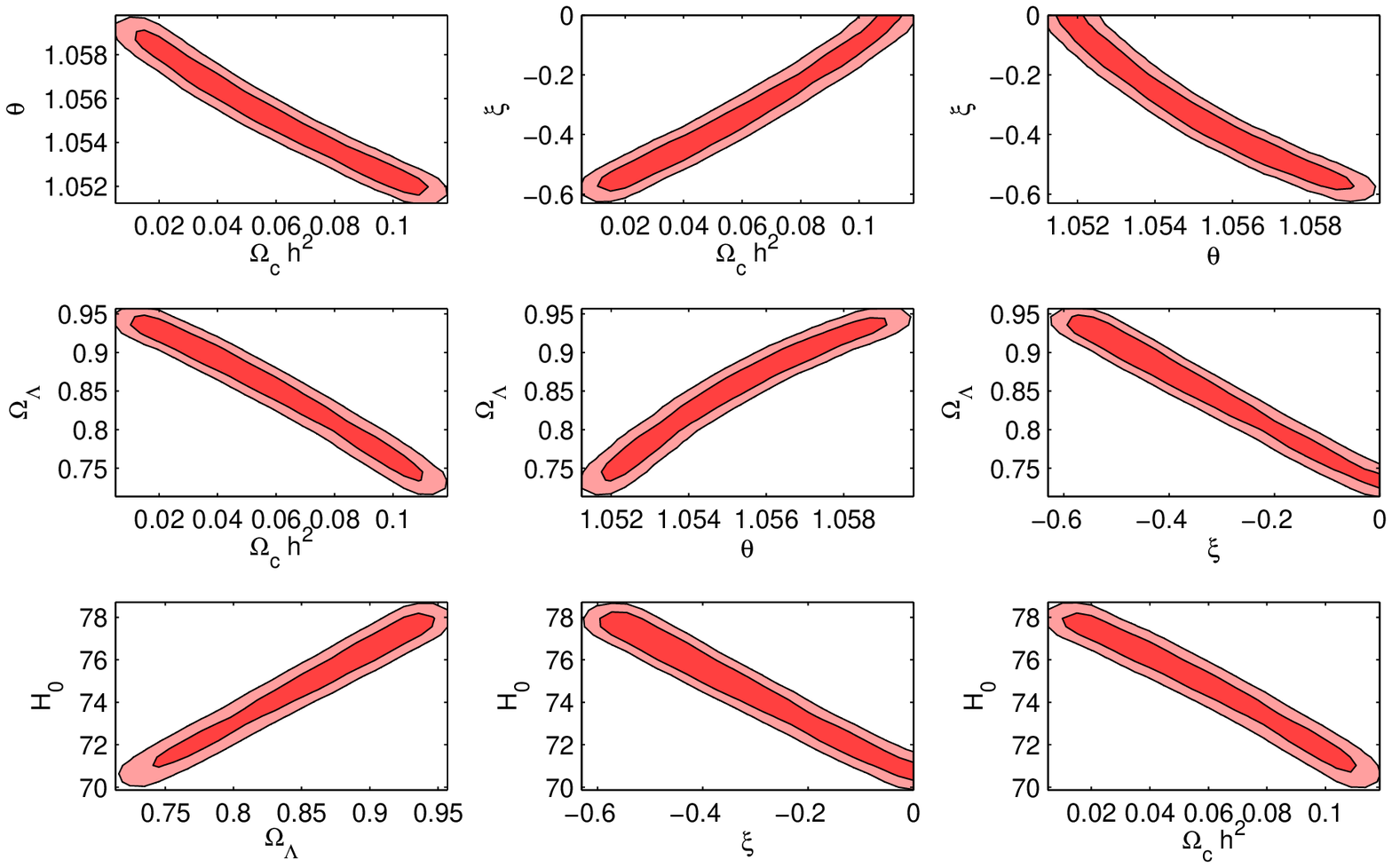}\\
 \end{tabular}
   \caption{The panels show the $68 \%$ and $95 \%$ confidence level contours 
combining the five most correlated parameters ($\Omega_ch^2$, $\theta_s$,
 $H_0$, $\Omega_\Lambda$ and $\xi$) arising from a fit to mock EPIC data  without including lensing extraction in the analysis.}
  \label{fig:degeneracyCMBno}
\end{center} 
\end{figure}

\begin{figure}[h!]
\begin{center}
\hspace*{-1cm}  
\begin{tabular}{cc}
\includegraphics[width=9cm]{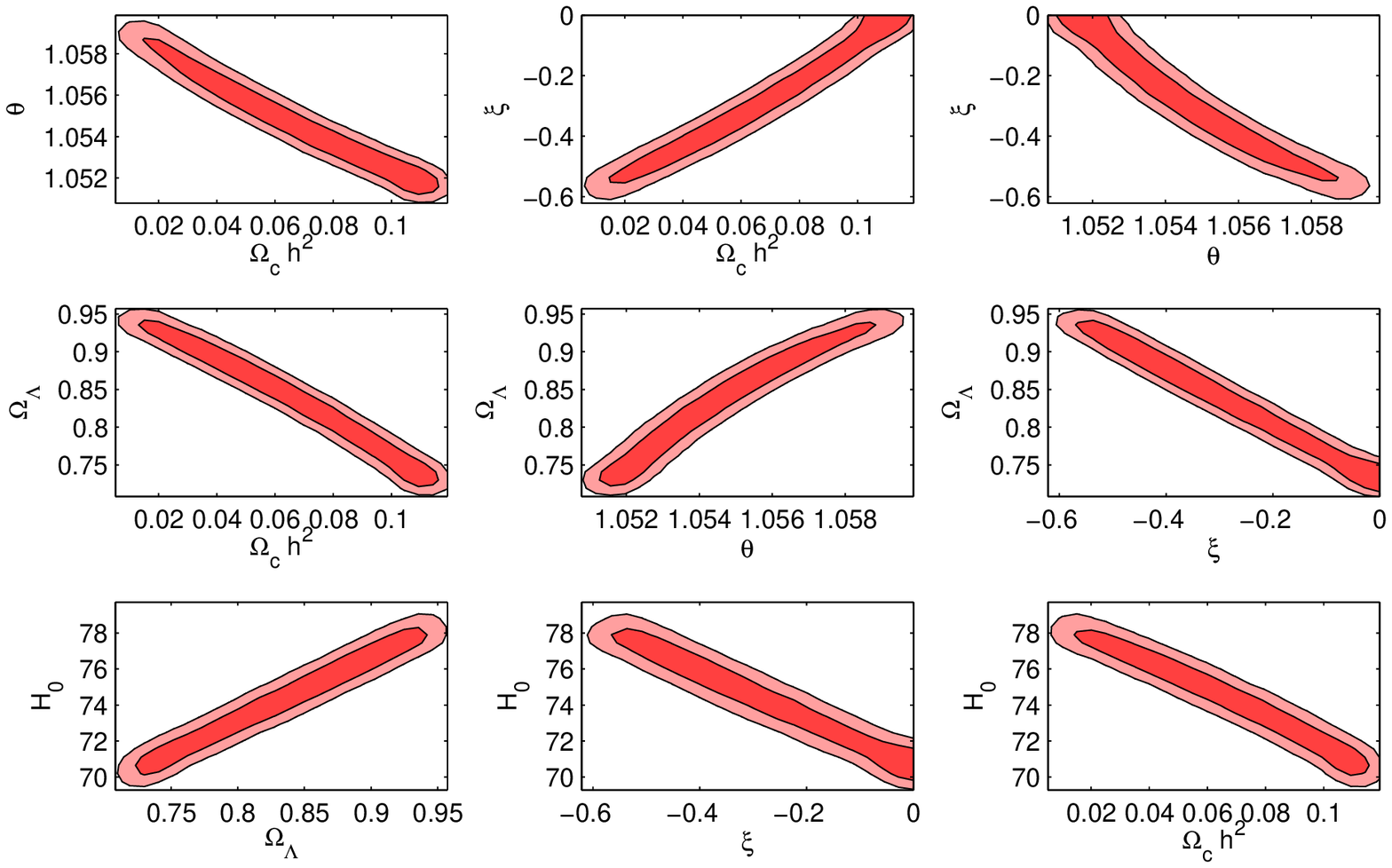}\\
 \end{tabular}
   \caption{The panels show the $68 \%$ and $95 \%$ confidence level contours 
combining the five most correlated parameters ($\Omega_ch^2$, $\theta_s$,
 $H_0$, $\Omega_\Lambda$ and $\xi$) arising from a fit to mock Planck data  including lensing extraction in the analysis.}

  \label{fig:degeneracyP}
\end{center} 
\end{figure}

\begin{figure}[h!]
\begin{center}
\hspace*{-1cm}  
\begin{tabular}{cc}
\includegraphics[width=9cm]{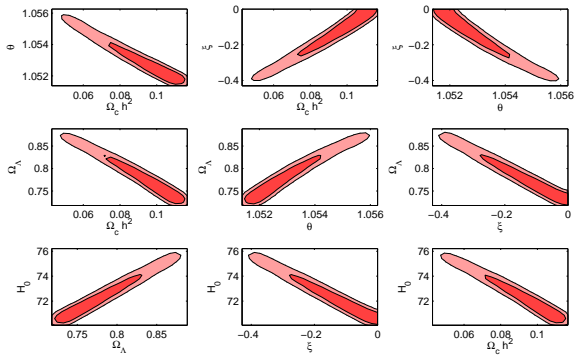}
 \end{tabular}
  \caption{The panels show the $68 \%$ and $95 \%$ confidence level contours 
combining the five most correlated parameters ($\Omega_ch^2$, $\theta_s$,
 $H_0$, $\Omega_\Lambda$ and $\xi$) arising from a fit to mock EPIC data including lensing extraction in the analysis.}
  \label{fig:degeneracyCMB}
\end{center} 
\end{figure}

Figures (\ref{fig:degeneracyPno}) and (\ref{fig:degeneracyCMBno}) depict the 
$68 \%$ and $95 \%$ confidence level contours combining the five most correlated parameters arising from a fit to mock Planck and EPIC data respectively, without considering CMB lensing extraction.
We can see that the cold dark matter density $\Omega_c h^2$, the Hubble constant $H_0$, the cosmological constant $\Omega_{\Lambda}$, the sound horizon angle $\theta_s$ and the coupling $\xi$ are all strongly correlated and neither Planck nor EPIC data will be able to break these degeneracies. Notice as well that, despite the technological advances of EPIC, the error on the cosmological parameters achieved by the Planck experiment will not be further improved by EPIC data if no lensing signal is considered.

Figures (\ref{fig:degeneracyP}) and (\ref{fig:degeneracyCMB}) show the 
$68 \%$ and $95 \%$ confidence level contours combining the five most correlated parameters arising from a fit to Planck and EPIC future data respectively, considering the information from CMB lensing. Notice that the inclusion of lensing power spectrum improves drastically the EPIC constraints. However, the addition of CMB lensing information does not change Planck results. This is due to the fact that the lensing noise for EPIC is significantly lower than for the Planck experiment, and therefore EPIC data would be able to reject models that otherwise would be accepted by Planck, see Fig.~(\ref{fig:lensnoise}).

\section{Conclusions}
\label{sec:seciiiii}
The current accelerated expansion of the universe is driven by the so-called dark energy. This negative pressure component could be interpreted as the vacuum energy density, or as a cosmic, dynamical scalar field.  If a cosmic quintessence field is present in nature, it may couple to the other fields in nature. While the couplings of the quintessence field to ordinary matter are severely constrained, an energy exchange among the dark matter and dark energy sectors is allowed by current observations. 

The major goals of the on-going Planck and the future EPIC experiments are to determine the nature of the dark energy component and to measure the remaining cosmological parameters with unprecedented precision. Several studies in the literature have been devoted to explore the performance of Planck and EPIC experiments in the dark energy scenario, see for instance Ref.~\cite{Perotto:2006rj}. In this paper we have explored the performance of Planck and EPIC experiments in alternative dark energy cosmologies, more concretely, in a universe with a coupling $\xi$ among the dark energy and dark mater components~\cite{Gavela:2009cy}.    

We have generated mock data for the Planck and EPIC experiments. CMB gravitational lensing extraction has also been included in the analysis. The lensing noise has been computed by means of the minimum variance estimator method of Ref.~\cite{lensextr}. The mock data have then been analyzed using MCMC techniques to compute the errors on the several cosmological parameters considered here.
We find that relevant degeneracies are present among the coupling $\xi$ and
 some other cosmological parameters, as the cold dark matter density $\Omega_{c}h^2$. Therefore, in the presence of a coupling, the expected Planck or EPIC errors on quantities as the cold dark matter energy density or the angular diameter distance at decoupling $\theta_s$ are one order of magnitude larger than in standard cosmologies with $\xi=0$. 

When gravitational CMB lensing extraction is included in the analysis, Planck results remain unchanged, due to the high level of lensing noise for this experiment. However, the EPIC mission, which will benefit from a much lower lensing noise level, can (a) provide tighter constraints on the cosmological parameters, even in the presence of a coupling, and (b) distinguish among coupled and uncoupled models that would look identical if they were fitted to Planck (lensed or unlensed) data. 

\section{Acknowledgments}
It is a pleasure to thank Belen Gavela for help and comments.
L. L. H. is as well supported by the PAU (Physics of the accelerating universe) Consolider Ingenio 2010, and by the FPA2009-09017 project. The work of O.M. is supported by a MICINN Ram\'on y Cajal contract, by AYA2008-03531 and the Consolider Ingenio-2010 project CSD2007-00060.

\end{document}